\documentclass[prb,superscriptaddress,aps]{revtex4}
\usepackage{graphicx}% Include figure files
\usepackage{dcolumn}% Align table columns on decimal point
\usepackage{bm}% bold math
\usepackage{amssymb}

\begin{document}

\title{On some quantum Hall states with negative flux}

\author{Th. Jolicoeur}
\affiliation{LPTMS, Universit\'e Paris-Sud, Building 100-A, 91405 Orsay}
\begin{abstract}
Some recently observed fractional quantum Hall states are not easily explained
in standard hierarchy/composite fermion schemes. This paper gives a brief introduction
to some wavefunctions involving non-Abelian Read-Rezayi states with negative flux
that have been proposed as candidates for these new quantum Hall fractions.

\textit{ Talk given at les Houches school ``Exact Methods in Low-dimensional Statistical Physics and Quantum Computing'',
Les Houches, France, June 30th - August 1st 2009.}
\end{abstract}
\maketitle

\section{introduction}

The fractional quantum Hall effect (FQHE) is a remarkable state of electronic matter
that occurs in two-dimensions and under a strong magnetic field. From a practical point of view
one observes experimentally a wealth of new thermodynamic phases in this special regime.
While there is a theory for the most prominent states (in a sense given below), recent experiments
have given evidence for more complexity for which there is yet no simple and universal theory.
This seminar gives an overview of the situation as of end of 2008 and focuses on some
recent proposals for wavefunctions describing these new quantum Hall states.

For the sake of completeness let us first set the stage for the FQHE. It is known to occur
for particles confined in two spatial dimensions and subjected to a perpendicular magnetic field.
In this set-up the one-body spectrum is drastically affected by the magnetic field~: indeed
kinetic energy is frozen and there is a set of exactly degenerate energy levels called Landau levels.
These levels are separated by the cyclotron energy $\hbar\omega_c$, $\omega_c =eB/m$.
Their degeneracy is given by $eB/h$ times the area $A$ of the sample. Increasing the $B$ field leads to
more degenerate levels and also to a larger separation between Landau levels. Imagine now
that we have some particles in this situation and we look for the ground state. If we have $N_e$ electrons
and we increase $B$, at some point there will be enough states in the lowest Landau level (LLL)
to accommodate everybody. This will happen first when $N_e =eB_1/h\times A$ hence for a density
$n=N_e/A=eB_1/h$. Beyond that value of $B$ there will be more states available and the problem
of putting the electrons in the one-body orbitals becomes exponentially degenerate since the number
of configurations is given by a binomial coefficient. If the cyclotron gap between the levels is large enough
it is the mutual interactions between electrons that will determine the structure of the ground state.
There is no longer any interplay between kinetic energy and potential energy. The FQHE
is a ``pure'' interaction effect. Typical two-dimensional electron gases have densities
of the order of $10^{11}$cm$^{-2}$. The field required to put everybody in the LLL is thus
$B_1\approx 4\times n(10^{11})$ Tesla. the number of occupied orbitals in the LLL is called
in what follows the filling factor $\nu$, the ratio of electrons divided by the number of one-body 
quantum states in the LLL $\nu = nh/eB$.

Let us gues what would have been the reasoning of a condensed matter physicist before the eighties
confronted with this situation. The exactly degenerate Landau levels are of course
an idealized situation and the disorder present in the real-world samples broadens
the levels which retains nevertheless a high density of states. These broadened levels are partially
filled for $B>B_1$ so, as a function of strength of interactions, there are two plausible
guesses~: the first one is a Fermi liquid in a partially filled band. Since the band is narrow
one can also envision the relevance of a Mott insulating state, i.e. a crystalline state of electrons.
This second possibility is in fact realized at very small filling factor, $\nu \lesssim 1/7$.
This is the so-called Wigner crystal. However the Fermi liquid phase does not happen
and is replaced by a new kind of liquid state supporting fractionalization of quantum numbers.
In fact there is a whole series of such liquids as a function of the filling factor $\nu$.
Historically the first state for which R. B. Laughlin~\cite{RBL} gave a satisfactory theoretical description
was $\nu =1/3$. The so-called ``Composite Fermion'' scheme developed by 
J. K. Jain~\cite{Jain89} 
gave a description of many other FQHE states. We will briefly give an overview of these
theoretical approaches in section II. 

Finally note that we will not discuss the transport properties of these FQHE states. It is certainly
true that the study of a FQHE state starts by a measurement of resistivity. However
the description of transport is a subject \textit{per se} 
and is described in several Les Houches lectures~\cite{LH}.
For us we just need to know that the FQHE states are liquids without any obvious local symmetry breaking,
they do possess a gapped excitation spectrum and amongst the excitations there are
unconventional quasiparticles with fractional charge and statistics. Experiments
are very often limited to gap measurements. The ``most prominent'' states are the ones
with largest gaps.

\section{classical hierarchies}
We now describe briefly the 
existing microscopic theory of the most prominent FQHE states~\cite{PrangeGirvin,DasSarma,Heinonen}.
If we use the symmetric gauge for the external applied magnetic field
$\mathbf{A}=\frac{1}{2}\mathbf{B}\times\mathbf{r}$, then a basis 
of the one-body eigenstates for the LLL is given by~:
\begin{equation}
 \phi_{m}(z)=\frac{1}{\sqrt{2\pi 2^m m!}} z^m {\mathrm e}^{-|z|^2/4\ell^2},
\label{eigs}
\end{equation} 
where $m$ is a positive integer, $\ell=\sqrt{\hbar/eB}$ is the magnetic length,
$z=x+iy$ is the complex coordinate in the plane.
In the absence of interactions and external potential all these states have exactly
the same eigenenergy $\frac{1}{2}\hbar\omega_c$. These states have a definite chirality
since $m$ is positive. The density associated with such a state is nonzero in the neighborhood
of a ring centered on the origin (due to the gauge choice) of radius $\ell\sqrt{2m}$.
We now consider the many-body problem with electrons all residing in the LLL.
The states we describe are also spin polarized i.e. the Zeeman energy is large enough.
This is not true in general of course and there are many interesting FQHE states involving
the spin degree of freedom but the states at very high field presumably do not involve the spin.
Each electron is described by a complex variable and thus the many-body wavefunction is of the form~:
\begin{equation}
 \Psi(z_1,\dots,z_N) = f(z_1,\dots,z_N)\,\, {\mathrm e}^{-\sum_i |z_i|^2/4\ell^2}.
\label{NbLLL}
\end{equation} 
In general, contemplation of the full many-body wavefunction is not a very useful way to understand
a physical system but in the FQHE physics this has proven to be the best approach (so far).
Let us first understand what happens when the LLL is full of electrons, $\nu =1$ or $B=B_1$
as defined in the introduction. We imagine a cylindrical infinite wall of a given radius $R$.
This has the effect to send to infinity all states in Eq.(\ref{eigs}) with $m$ larger than
$R^2/2\ell^2$. Only the states close to the boundary will have wavefunctions different from
the formula Eq.(\ref{eigs}). This is a negligible effect for large systems. The number of states
in the LLL is thus finite and equal to $M=R^2/2\ell^2=eB/h\times$(area of cylinder). If we fill
exactly all these one-body states $\propto z^0,z^1,\dots,z^M$
with fermions there is a unique possible state for $B=B_1$ which is
simply the Slater determinant of all occupied orbitals~:
\begin{equation}
 \Psi_{\nu =1}= {\mathrm Det}\left[ z_i^{j-1}\right] \,\, {\mathrm e}^{-\sum_i |z_i|^2/4\ell^2},
\end{equation} 
where indices $i,j$ run from 1 to $N$. This determinant is called the Vandermonde determinant
and can be computed in closed form~: 
${\mathrm Det}\left[ z_i^{j-1}\right]=\prod_{i<j}(z_i-z_j)$. While this leads to a nice
wavefunction this is not a very fascinating one. Since all orbitals are occupied 
this is the exact ground state for any interactions between electrons if we neglect the possibility
of transition towards higher lying LLs (note that when there is spin in the game things are now nontrivial -
we will not consider this important point). The density distribution of $\psi_{\nu=1}$
can be guessed easily~: since all orbitals are occupied in a uniform manner and these orbitals fill
concentric rings of increasing radius with the same peak density,
we have uniform coverage of a circular region. The state looks like a flat pancake which decays to zero
on a length scale given by $\ell$ only close to the boundary. Encouraged by this understanding
let us try to guess a candidate ground state for a filling factor less than unity. This was the approach used 
by Laughlin. He proposed to take simply the cube of the Vandermonde determinant. Since all factors
$(z_i-z_j)$ becomes $(z_i-z_j)^3$, this repels the electrons and flattens the pancake of the charge distribution.
It is easy to guess that the new wavefunction is still pancake-like but with now a density which is 1/3
of the previous $\nu=1$ case hence it is a state with $\nu=1/3$~:
\begin{equation}
 \Psi_{\nu=1/3}=\prod_{i<j}(z_i-z_j)^3 \,\, {\mathrm e}^{-\sum_i |z_i|^2/4\ell^2}.
\end{equation} 
Why is this considered as being close to the truth for interacting electrons at $\nu =1/3$ ? Contrary
to the case of $\nu =1$ this is \textit{not} an exact eigenstate of the Coulomb problem.
We first need to make a \textit{d\'etour} by looking at the two-body problem in the LLL.
If we consider the kinetic energies of two charged particles then there is the following identity~:
\begin{equation}
 \frac{1}{2m}(\mathbf{p}_1+e\mathbf{A}_1)^2 +
\frac{1}{2m}(\mathbf{p}_2+e\mathbf{A}_2)^2 =
\frac{1}{2M}(\mathbf{P}_{cm}+2e\mathbf{A}_{cm})^2 +
\frac{1}{2\mu}(\mathbf{p}_r+\frac{e}{2}\mathbf{A}_r)^2 .
\end{equation} 
where we have separation of the center of mass $M=2m$ and the relative particle motions.
The relative particle also is living in Landau levels so its kinetic energy is frozen.
To find the eigenenergies of the two-body problem is now trivial~: we just have
to take expectation values of the interaction potential $V(\mathbf{r}_1-\mathbf{r}_2)$
in the eigenstates of the relative particles given by eq.(\ref{eigs}) with $8\ell^2$
in the exponential instead of $4\ell^2$ since $e/2$ appears in the relative kinetic energy.
These eigenenergies are thus $V_m\equiv\langle\phi_m|V|\phi_m\rangle$ for any (rotationally invariant!)
potential (forgetting the cyclotron energy independent of $m$). 
The exponent $m$ appearing in the relative particle wavefunction is positive,
i.e. the relative angular momentum is always positive. So any two-body interaction in the LLL
is parametrized fully by the $V_m$ coefficients called the pseudopotentials 
after D. Haldane~\cite{PrangeGirvin}.
This peculiarity of the two-body problem also means that one can write
the interaction Hamiltonian in a very special way~:
\begin{equation}
 \mathcal{H}=\sum_{i<j}\sum_{m}V_m\mathcal{P}^m_{ij},
\end{equation} 
where $\mathcal{P}^m_{ij}$ projects the couple of particles $i,j$ onto
relative angular momentum $m$ and the sum over $m$ is restricted to positive
odd integers for spinless fermions due to Pauli statistics. 
For the repulsive Coulomb potential these pseudopotentials
decrease with increasing values of $m$. There is a very fundamental property of the Laughlin
wavefunction~: it is the unique smallest degree homogeneous polynomial which is a zero-energy
eigenvalue of the special model where only $V_1$ is non-zero. This allows to understand why we believe
that the Laughlin wavefunction captures the correct physics. When we vary pseudopotentials
between the hard-core model with only non-zero $V_1$ and the true Coulomb problem, there is
clear numerical evidence that nothing dramatic happens, i.e. we do not cross any phase
boundary. This has been shown by D. Haldane by exact diagonalization of small systems~\cite{PrangeGirvin}.

The Laughlin wavefunction can only describe liquids with filling factors $1/m$, $m$ odd
while there are many more FQHE states in the real world. We start to trying to rewrite
the Laughlin state as a determinant. We like determinant since Slater determinants
are the simplest way to get an efficient description of atoms, molecules, nuclei and solids.
With Slater determinants we can make particle-hole excitations and hence construct excited states
on top of the ground state. This is a very desirable theoretical tool.
We do the following manipulation~:
\begin{equation}
 \Psi_{\nu=1/3}= \prod_{i\neq j}(z_i-z_j)\times \prod_{i<j}(z_i-z_j),
\label{redflux}
\end{equation} 
where the ubiquitous exponential factor is not written for clarity. The last factor
is the Vandermonde determinant. We note that~:
\begin{equation}
 \prod_{i\neq j}(z_i-z_j)=\prod_{j\neq 1}(z_1-z_j) \dots  \prod_{j\neq N}(z_N-z_j)
\equiv J_1 \dots J_N,
\end{equation} 
where we have defined the so-called Jastrow factors $J_i$. These factors can be distributed
along the columns of the Vandermonde determinant~:
\begin{equation}
  \Psi_{\nu=1/3}=J_1 \dots J_N\times \prod_{i<j}(z_i-z_j)=
\left|
\begin{array}{ccc}
J_1 & \dots & J_N \\
z_1 J_1 & \dots & z_N J_N \\
: & : & : \\
: & : & : \\
z_1^{N-1}J_1 & \dots & z_N^{N-1}J_N 
\end{array}
\right|.
\end{equation} 
So this \textit{is} a Slater determinant provided we change the rules in the following way~:
instead of using orbitals $z^m$ which are bona fide one-body orbitals
we now use pseudo-orbitals $z^m J$ where $J$ effectively repels all the other particles.
This is not really a one-body object but within this construct we can play the same usual
Slater-like construction of excited states and so on.
The first appearance of the Vandermonde determinant in Eq.(\ref{redflux})
is suggestive of a flux reduction effect of correlations. Indeed the Vandermonde determinant
is the ground state at $\nu =1$. It is as if the correlation factors $J_1\dots J_N$ reduce
the magnetic field from $B$ to $B_{eff}=B-2nh/e$ so that $\nu =1/3$ becomes $\nu =1$.
If we have~:
\begin{equation}
 \Phi_{\nu}= \prod_{i<j}(z_i-z_j)^2 \Phi_{\nu^{*}},
\end{equation} 
then
the two filling factors are related by $1/\nu = 2+ 1/\nu^*$. This is intuitively reasonable~:
the Jastrow factor repels the particles and flattens the pancake i.e. the charge distribution
of the wavefunction. Let us now reason in terms of the effective filling factor
$\nu^*$. If $\nu$ is greater than 1/3 it implies that the effective $\nu^*$ is now larger
than one. We thus have to occupy Landau levels higher than the LLL in the Slater determinant.
There is nothing wrong with that, provided we project back to the LLL. There will be special
filling when there is filling of an integer number $p$ of LLs. It is natural but not immediately
obvious to expect that such wavefunctions will have to do with incompressible FQHE states.
The candidate ``composite fermion'' (CF) states~\cite{Jain89,Heinonen} are thus~:
\begin{equation}
\Phi_{\nu =p/(2p+1)}= {\mathcal P}_{LLL} \prod_{i<j}(z_i-z_j)^2 \Phi_{\nu^{*}=p}
\label{CF1}
\end{equation}  
For example the case $p=2$ involves the second LL which is spanned by the one-body
orbitals $z^* z^m$ with $m\geq 0$. We can make a Slater determinant by putting half of the electrons 
in the pseudo LLL and the other half in the second pseudo-LL~:
\begin{equation}
 \Phi_{\nu^{*}=2}=
\left|
\begin{array}{ccc}
1 & \dots & 1 \\
z_1 & \dots & z_N \\
: & : & : \\
: & : & : \\
z_1^{N/2-1} & \dots & z_N^{N/2-1} \\
z^*_1 & \dots & z^*_N \\
z^*_1 z_1 & \dots & z^*_N z_N \\
 : & : & : \\
: & : & : \\
z^*_1 z_1^{N/2-1} & \dots & z^*_N z_N^{N/2-1}
\end{array}
\right|.
\end{equation} 
This is essentially the original Jain proposal - it has proven extremely successful.
However manipulation of this wavefunction is inconvenient in practice due to the
projection that one has to perform after multiplication by the Jastrow factor.
Jain and Kamilla have shown that this scheme may be slightly altered to be much more tractable while retaining 
all its good quantitative properties. The idea is again to distribute all $J$ factors
in the determinant as before but now we project \textit{before} computing the determinant.
It means that we just have to project onto the LLL factors like 
$z^*z^m J$. This is done by replacing each $z^*$ by $\partial/\partial z$.
Doing the derivations is now trivial and after some elementary manipulations we have~:
\begin{equation}
 \Phi_{\nu =2/5}=  \prod_{i<j}(z_i-z_j)^2 
\left|
\begin{array}{ccc}
1 & \dots & 1 \\
z_1 & \dots & z_N \\
: & : & : \\
: & : & : \\
z_1^{N/2-1} & \dots & z_N^{N/2-1} \\
\Sigma_1 & \dots & \Sigma_N \\
 z_1 \Sigma_1& \dots & z_N \Sigma_N\\
 : & : & : \\
: & : & : \\
 z_1^{N/2-1} \Sigma_1 & \dots & z_N^{N/2-1}\Sigma_N
\end{array}
\right|,
\end{equation} 
where $\Sigma_i =\sum_{j\neq i}\frac{1}{z_i-z_j}$.
This is a typical example of the CF scheme. Note that one can
change the partitioning between the two LLs involved by putting
say $N_1$ electrons in the LLL and $N_2$ in the second LL with $N_1+N_2=N$.
All these states are observed in exact diagonalizations in the unbounded
disk geometry. The reason why we believe in the CF is slightly different
wrt the Laughlin state. Here there is no simple Hamiltonian
for which such states are unique exact ground states. But the spectroscopy
of low-lying levels is correctly reproduced~: we find the good quantum 
numbers and ordering of levels in the CF scheme when compared to exact
diagonalization results. This CF scheme gives a reasonable description
of FQHE states at $\nu=p/(2p+1)$. By multiplying by extra powers of the Vandermonde
it is easy to generate candidates for $p/(4p+1)$, $p/(6p+1)$ ...
without new ideas. In the CF folklore we say that CF are electrons dressed by two flux tubes
which means that there is a Jastrow factor squared that appears in the trial wavefunction.

Note now that for $B$ less than $2nh/e$ the effective field is now \textit{negative}.
This happens for fractions between $\nu=1$ and the limiting case $\nu=1/2$ (which is a compressible
state). This is easily included in the CF scheme since changing the sign of $B$
amounts to complex conjugation. We just have to use $p$ filled CF levels
with negative fluxes in Eq.(\ref{CF1}). While there are more derivative operators,
it also leads to satisfactory wavefunctions describing fractions now at 
$p/(2p-1)$, $p/(4p-1)$ and so on.
This CF construction also explains naturally the occurrence of a gapless compressible state
at $\nu =1/2$~: this is the value for which the net effective flux is zero, suggestive
of freely moving CF particles.
For many years it has been known that one fraction with $\nu =5/2$ does not fit into
the CF scheme. This FQHE state is in the second LL and is in fact a state with
partial filling 1/2 of the second LL. It is reasonable to expect that the filled
LLL with the two spin values i.e. $\nu =2$ plays the role of an inert dielectric medium
renormalizing the interactions hence we should find a replica of the FQHE phenomenon
in the second LL if the interactions are not dramatically altered. In fact
there is a big difference which is the appearance of a FQHE at $\nu=5/2$ with an \textit{odd} denominator,
forbidden from CF/hierarchical constructions. The most successfull candidate
to describe this state is the Moore-Read Pfaffian state~\cite{Moore91,Greiter92}
which is a state
with pairing between the effective CF particles. This state has fractionally charged
excitations with charge $e/4$ for which there is some experimental evidence.
It is described below.

Recent experiments~\cite{Pan03} have uncovered states displaying 
FQHE  at filling factors
$\nu =4/11, 5/13, 4/13, 6/17$, and $5/17$ that do not belong to the primary
FQHE sequences. In addition, there is also evidence for two new \textit{even}-denominator fractions
 $\nu =3/10$, and $3/8$. This is very unusual since the only previously known example
of an even-denominator fraction is the elusive $\nu =5/2$ state. The state $3/8$ has 
also been observed~\cite{Xia04} in the N=1 LL at total filling factor $\nu = 2+3/8$.
The new odd-denominator fractions can be explained by hierarchical reasoning
in the spirit of the original Halperin-Haldane hierarchy. For example, at $\nu =4/11$, the 
CFs have an effective filling factor $\nu_{CF} = 1+ 1/3$. 
If the interactions between the CFs
have a repulsive short-range core then it is plausible that they will themselves form a
standard Laughlin liquid at filling factor 1/3 within the second CF Landau level.
It should be pointed out that this construction of
``second generation'' of composite fermions is part of the 
standard lore of the
hierarchical view of the FQHE states since the CF construction and the older Halperin-Haldane
hierarchy can be related by a change of basis in the lattice of quantum numbers~\cite{Read90}.
Since the even-denominator fractions requires clustering they do not fit naturally in this picture.
There is no natural ``Grand Unification'' of all these new fractions in the hierarchical constructs.

It is possible~\cite{thj} to make a construction based on the idea of composite bosons that carry now
an \textit{odd} number of flux quanta i.e. a Jastrow factor to an odd power
in the trial wavefunction.
This has to be contrasted with the previous CF construction where
we used only even powers of the Jastrow factor.
These fluxes may be positive or negative. One can then exploit the 
possibility of clustering of bosons in the lowest LL (LLL). Indeed it has been suggested~\cite{Cooper01} 
that incompressible liquids of Bose particle may form at fillings $\nu =k/2$ with integer $k$. 
We will now write down
spin-polarized FQHE wavefunctions on the disk and spherical geometry. By construction they reside
entirely in the LLL and have filling factor $\nu = k/(3k\pm 2)$. 
While the positive flux series
 already appeared in
the work of Read and Rezayi~\cite{Read96,Read99}, the negative flux series is new. 
These series produce candidate wavefunctions for all the states observed by Pan et al. beyond
the main CF sequences, thus unifying even and odd denominator fractions.
For the fraction 3/7, the negative flux
candidate wavefunction has an excellent overlap with the Coulomb ground state obtained by exact
diagonalization on the sphere for N=6 electrons. 

The first observation is that some of the new fractions of ref.(\cite{Pan03}) are of the form
$p/(3p\pm 1)$. This would be natural for the FQHE of \textit{bosons} where one 
expects the formation
of composite fermions with an \textit{odd} number of flux tubes, i.e. $^1$CF and $^3$CF. 
The $^1$CF lead to a series of Bose fractions at $\nu = p/(p+1)$ which has nothing to do with
the present problem.
But if the $^3$CFs fill an integer number of pseudo-Landau levels then this leads
to magic filling factors $p/(3p\pm 1)$. Indeed there is evidence
from theoretical studies of bosons in the LLL with dipolar interactions~\cite{Rezayi05}
 that such $^3$CF do appear.
This suggests that composite bosons may form in the electronic system, three flux tubes bound 
to one electron, $^3$CBs, the attachment may be with statistical flux along or against the 
applied magnetic field. If $\nu$ stands for the electron filling factor and $\nu^*$ the 
$^3$CB filling factor, 
they are related by $1/\nu = 3 +1/\nu^*$. The relationship 
between the wanted electronic trial wavefunction and the CB wavefunction is~:
\begin{equation}
\Psi^{Fermi}_\nu (\{z_i\})=
\mathcal{P}_{LLL}\prod_{i<j}(z_i-z_j)^3\,\, \Phi^{Bose}_{\nu^*}(\{z_i,z_i^*\}).
 \label{wavef}
\end{equation} 
The Laughlin-Jastrow factor $\prod_{i<j}(z_i-z_i)^3$ transforms bosons into fermions and
adequately takes into account the Coulomb repulsion.
The next step is to find candidates for the trial state $\Phi^{Bose}_{\nu^*}$.
It has been suggested~\cite{Cooper01} that bosons in the LLL may form incompressible
states for $\nu^*=k/2$. There is some evidence that they are described by
 the Read-Rezayi parafermionic states~\cite{Read96,Read99} with clustering of $k$
particles~:
\begin{equation}
 \Phi^{RR}_{\nu^*=k/2}=\mathcal{S}\,\,
\left[
\prod_{i_1<j_1}(z_{i_1}-z_{j_1})^2 \dots
\prod_{i_k<j_k}(z_{i_k}-z_{j_k})^2 
\right].
\label{RR}
\end{equation} 
In this equation, the $\mathcal{S}$ symbol means  symmetrization of the product
of Laughlin-Jastrow factors over all partition of N particles in subsets of $N/k$
particles ($N$ being divisible by $k$)
(the ubiquitous exponential factor appearing in all LLL states has been omitted for clarity).
While the relevance of such states to bosons
with contact interactions is not clear, it has been shown that longer-range interactions 
like dipolar
interaction may help stabilize these states~\cite{Rezayi05}. Since the CBs are composite objects
it is likely that their mutual interaction has also some long-range character. It is
thus natural to try the ansatz $\Phi^{Bose}_{\nu^*}=\Phi^{RR}_{\nu^*=k/2}$ in Eq.(\ref{wavef}).
This leads to a series of states with electron filling factor $\nu =k/(3k+2)$ which is in fact
the $M=3$ case of the generalized $(k,M)$ states constructed by Read and Rezayi. In this construction, 
the flux attached to the boson is positive. It is also natural to construct wavefunctions
with negative flux~\cite{Moller05} attached to the CBs. Now the Bose function depends only upon
the antiholomorphic coordinates~:
\begin{equation}
\Phi^{Bose}_{\nu^*}(\{z_i^*\})=(\Phi^{RR}_{\nu^*=k/2}(\{z_i\}))^*
 \label{negflux}
\end{equation} 
The projection onto the LLL in Eq.(\ref{wavef}) means that the electronic wavefunction
can be written as~:
\begin{equation}
\Psi^{Fermi}_\nu (\{z_i\})= \Phi^{RR}_{\nu^*=k/2}(\{\frac{\partial}{\partial z_i}\})
\prod_{i<j}(z_i-z_j)^3 .
 \label{NGwavef}
\end{equation} 
The filling factor of this new series of states is now $\nu = k/(3k-2)$.
These states can be written in the spherical geometry with the help of the spinor components
$u_i=\cos (\theta_i/2) {\rm e}^{i\varphi_i/2}$,
$v_i=\sin (\theta_i/2) {\rm e}^{-i\varphi_i/2}$ ($\{\theta_i,\phi_i\} $ being standard polar coordinates) 
by making the following substitutions~:
\begin{equation}
z_i-z_j \rightarrow u_iv_j - u_j v_i ,
\quad \partial_{z_i}-\partial_{z_j}
\rightarrow 
\partial_{u_i}\partial_{v_j} -
\partial_{v_i}\partial_{u_j}.
\end{equation} 
This construction leads to wavefunctions that have zero total angular momentum $L=0$
as expected for liquid states.
On the sphere these two series of states have a definite relation between the number
of flux quanta through the surface and the number of electrons. The positive flux series
has $N_\phi = N/\nu - 5$ while the negative flux series has $N_\phi = N/\nu - 1$.
Even when these states have the same filling factor as standard hierarchy/composite fermion states,
the shift (the constant term in the $N_\phi - N$ relation) is in general different.
The positive flux series starts with the Laughlin state for $\nu =1/5$ at $k=1$,
the $k=2$ state is the known Pfaffian state~\cite{Moore91,Greiter92} at $\nu=1/4$,
at $k=3$ there is a state with $\nu =3/11$ which competes with the $^4$CF state with negative flux,
at $k=4$ the competition is with the similar $\nu =2/7$  $^4$CF state. This series also contains
5/17 at $k=5$, 3/10 at $k=6$, and 4/13 at $k=8$. The negative flux series starts with the filled
Landau level at $k=1$ and contains notably 5/13 ($k=5$), 3/8 ($k=6$), 4/11 ($k=8$), 6/17 ($k=12$).
It is not likely that these states will compete favorably with the main sequence CF states
in view of the remarkable stability of the latter. However the situation is open concerning the exotic even
 denominator and the unconventional odd-denominator states. Also the CF states may be destabilized
by slightly tuning the interaction potential. 
% A two-body interaction in a given LL may always
% be parameterized by the pseudopotentials $V_m$, $m=1,3,\dots $,where $V_m$ is the 
% interaction energy for a single pair of electrons with relative angular momentum $m$
% (all energies will be expressed in units of $e^2 /\epsilon l_0 $ and $l_0=\sqrt{\hbar c/eB} $).
It is known for example that there is a window of stability
for a non-Abelian $\nu =2/5$ state in the N=1 LL~\cite{Rezayi06} which is obtained by
slightly decreasing the $V_1$ pseudopotential component with respect to its Coulomb value.

%%%%%%%%%%%%%%%%%%%%%%%%%%%%%%%%%%%%%%%%%%%%%%%%%%%%%%%%%%%%%%%%%%%%%%%%%%%%%%%
% the 3/7 story
%%%%%%%%%%%%%%%%%%%%%%%%%%%%%%%%%%%%%%%%%%%%%%%%%%%%%%%%%%%%%%%%%%%%%%%%%%%%%%%
\begin{figure}
  \begin{center}
    \begin{tabular}{cc}
      \resizebox{65mm}{!}{\includegraphics{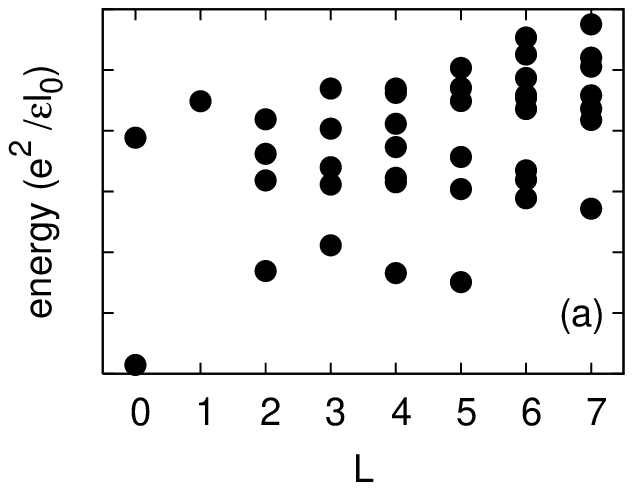}} &
      \resizebox{65mm}{!}{\includegraphics{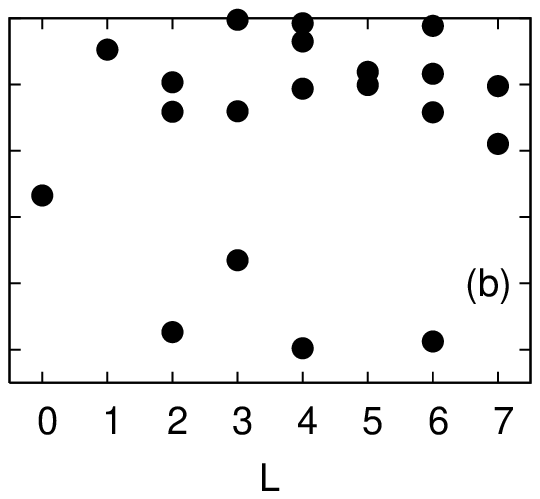}} \\
      \resizebox{65mm}{!}{\includegraphics{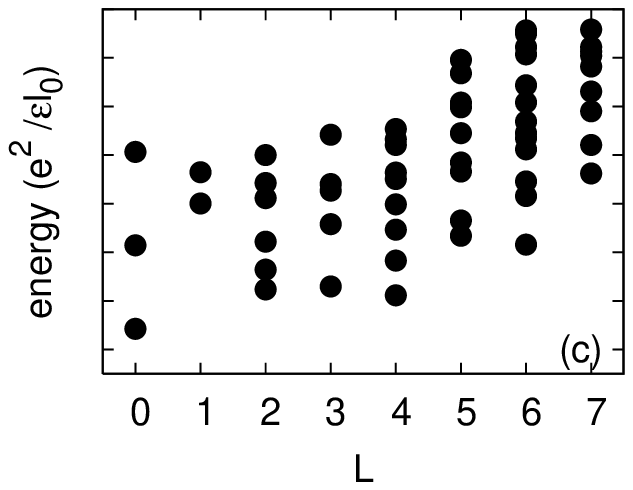}} &
      \resizebox{65mm}{!}{\includegraphics{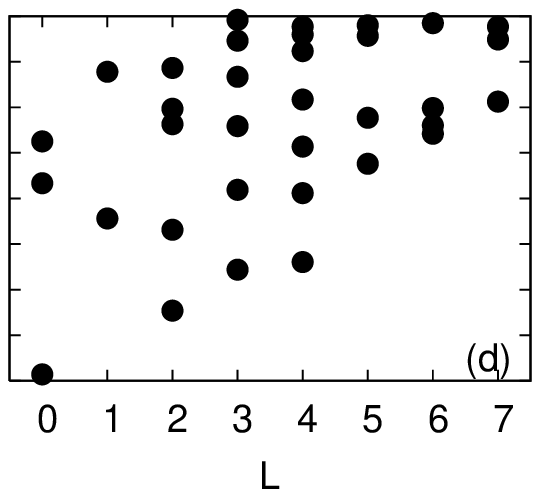}} \\
    \end{tabular}
    \caption{Low-lying spectrum of 9 electrons in the LLL
     as a function of the angular momentum
     on a sphere. 
     Top panel~: Coulomb interaction.
     (a) At
     $N_\phi=16$, the flux needed for the 3/7 CF state there is a singlet ground state
     and a branch of collective excitations. (b) At the flux needed for the candidate state
     there is no evidence of a FQHE state.
     Bottom panel weakened potential with $V_1 = 0.7V_1^{Coulomb}$~:
      (c) The 3/7 CF state is now compressible.
      (d) There is a possible new FQHE state
     with the shift required by Eq(\ref{NGwavef}).}
    \label{coulomb37}
  \end{center}
\end{figure}

\begin{figure}
    \begin{center}
      \resizebox{70mm}{!}{\includegraphics{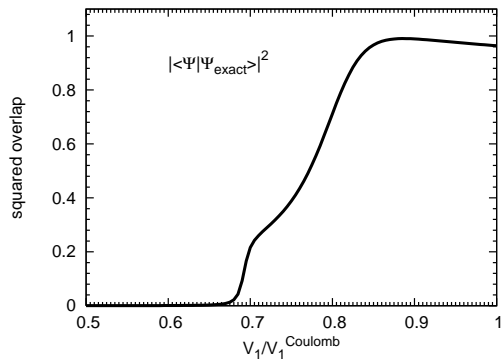}}
      \caption{The squared overlap  for N=6 electrons at $N_\phi=13$ between the candidate
        wavefunction at $\nu =3/7$
        and the exact ground state computed by varying the pseudopotential $V_1$ with respect
         to the Coulomb value.}
      \label{overlap}
    \end{center}
  \end{figure}

A similar phenomenon seems to happens at $\nu =3/7$ in the LLL. The conventional CF
state at this filling factor is a member of the principal sequence of states. It
is realized for $N=9$ electrons at $N_\phi =16$ in the spherical geometry. There is a 
singlet ground state and a well-defined branch of neutral excitations
for $L=2,3,4,5$~: see Fig.(\ref{coulomb37}a).
 The negative-flux state Eq.(\ref{NGwavef})
requires $N_\phi=20$ for the same number of particles. At this flux for pure Coulomb interaction
there is simply a set of nearly degenerate states without evidence for an incompressible 
state~: see Fig.(\ref{coulomb37}b). If the pseudopotential $V_1$ is decreased from its Coulomb
LLL value, the CF state is quickly destroyed (Fig.(\ref{coulomb37}c)) but there is appearance of a possibly
incompressible state precisely at the special shift predicted above~: Fig.(\ref{coulomb37}d). 
There is a $L=0$ ground state and
a branch of excited states for $L=2,3,4$. To check if this state has anything to do with
the new negative flux state
proposed above,  the overlap
between the candidate wavefunction for $k=3$ in Eq.(\ref{NGwavef}) and the numerically 
obtained ground state
is displayed in Fig.(\ref{overlap}) for N=6 electrons at $N_\phi=13$. Even for the pure 
Coulomb interaction
the squared overlap is 0.9641 and it rises up to 0.99054 for $V_1=0.885V_1^{Coulomb}$.

More numerical evidences may be found in ref.(\cite{thj}) concerning some
of the other fractions in these series like 3/8 and 3/11.
It is difficult to study states with high-order $k$-clustering since they require
at least $2k$ particles. It should be noted that so far these new series of states
have not been derived from correlators of a conformal field theory. It is known
that the Read-Rezayi states can all be derived from expectation values
of fields of parafermionic CFTs. Proving that the new states we described above can be derived from
a unitary CFT would be an indication that they describe incompressible candidate
FQHE states~\cite{Read09}.

%\acknowledgements

\end{document}